\documentclass[12pt]{article}
\usepackage{amssymb}
\usepackage{graphicx}
\def\beq{\begin{equation}}
\def\eeq{\end{equation}}
\usepackage[all,2cell,dvips]{xy}
\newtheorem{proposicion}{Proposition}
\newtheorem{demosprop}{Proof of Proposition}

\def\IR{\relax{\rm I\kern -.18em R}}
\begin{document}
\title{A new integrable equation valued on a Cayley-Dickson algebra}
\author{\Large A. Restuccia*, A. Sotomayor**, J. P. Veiro***}
\maketitle{\centerline{*Department of Physics}}
\maketitle{\centerline{Universidad de Antofagasta, Chile}
\maketitle{\centerline {**Department of Mathematics,}}
\maketitle{\centerline{Universidad de Antofagasta, Chile}}
\maketitle{\centerline {***Department of Mathematics,}}
\maketitle{\centerline{Universidad Sim\'on Bol\'{\i}var, Caracas, Venezuela}}
\maketitle{\centerline{e-mails: alvaro.restuccia@uantof.cl,
adrian.sotomayor@uantof.cl, jpveiro@usb.ve }}
\begin{abstract}We introduce a new integrable equation valued on a Cayley-Dickson (C-D) algebra. In the particular case in which the algebra reduces to the complex one the new interacting term in the equation cancells and the equation becomes the known Korteweg-de Vries equation. For each C-D algebra the equation has an infinite sequence of local conserved quantities. We obtain a B\"{a}cklund transformation in the sense of Walhquist-Estabrook for the equation for any Cayley-Dickson algebra, and relate it to a generalized Gardner equation. From it, the infinite sequence of conserved quantities follows directly. We give the explicit expression for the first few of them. From the B\"{a}cklund transformation we get the Lax pair and the one-soliton and two-soliton solutions generalizing the known solutions for the quaternion valued KdV equation. From the Gardner equation we obtain the generalized modified KdV equation which also has an infinite sequence of conserved quantities. The new integrable equation is preserved under a subgroup of the automorphisms of the C-D algebra. In the particular case of the algebra of octonions, the equation is invariant under $SU(3)$.

\end{abstract}


Pacs: 02.30.lk, 11.30.-j, 02.30. Jr, 02.10.Hh

\section{Introduction}Since the work of Miura and Gardner et al \cite{Miura1,Miura2,Miura3,Miura4,Miura5},
 carried out in the context of the Korteweg-de Vries (KdV) equation, a lot of interesting results about the so-called integrable systems
 have been obtained. The fascinating theory developed from that time involves, among other features, several methods to solve non-linear partial differential equations which arise in a wide range of mathematical and physical situations. This is the case, for example, of the Hirota bilinear method \cite{Hirota1}, which allows to obtain multisolitonic solutions and of the B\"{a}cklund transformation obtained by Wahlquist and Estabrook \cite{Wahlquist} with its corresponding generating formula of solutions (including multisolitonic solutions) which acts as a nonlinear superposition principle.

One important aspect of these systems is the existence of infinite conserved quantities generalizing the well known property of totally integrable systems describing a finite number of degrees of freedom. See, for example \cite{Sepe}, for a review of these systems.  Also  the quantum formulation of such systems and its classical limit obtained from the commuting conserved charges is an interesting topic.

A nice idea due to Gardner \cite{Miura1} allows to obtain the infinite conserved quantities for the KdV equation using simultaneusly a parametric auxiliar equation (the Gardner equation) and a parametric transformation (the Gardner transformation), relating solutions between both equations. 

In \cite{Chen} Chen obtained the Wahlquist-Estabrook formulation for the KdV equation starting from the associated Lax equations and using a discrete symmetry of the associated Gardner equation (see also \cite{Olver}). Satsuma in \cite{Satsuma} used the Wahlquist-Estabrook transformation to get the corresponding infinite sequence of conserved quantities.

Following these ideas, it is a natural question to ask which of the possible extensions of the KdV equation still possess the integrability properties of the KdV equation. An integrable Grassmann valued extension of KdV was obtained in \cite{Kupershmidt}. A
class of extensions of KdV equation arises by introducing supersymmetry. Several integrable supersymmetric extensions were given in \cite{Mathieu1,Mathieu2,Bellucci,Delduc1,Delduc2,Popowicz}. These
extensions are a special subset of the set of the coupled
extensions. Coupled extensions of the KdV equation form a category by itself,
containing among others the ones giving in
\cite{Hirota2,Ito,Sakovich}. In particular, in \cite{Adrian10,Adrian11} we considered a $\lambda$-coupled KdV system. The case $\lambda=0$ plays a relevant role in the description of 3-dimensional gravity \cite{Troncoso1}.

Interesting extensions of the KdV equation follow by considering the
defining field to be valued either in associative algebras \cite{Marchenko}, see also
\cite{Sokolov,Svinopulov}, or in non-associative
algebras arising as non-commutative generalizations
of Jordan ones \cite{Sokolov1}. The complex KdV equation is also integrable with interesting blow-up properties \cite{Birnir}.

Another extensions of KdV can be given by considering a supersymmetric extension and then substituting the defining fields by Clifford valued fields \cite{Adrian4,Adrian5} (in order to analize a supersymmetric breaking procedure with nice stability properties for the resulting solitonic solutions) or by operators acting in some general space of functions \cite{Adrian1}. 

In the present work we introduce a new integrable equation with the field valued on a Cayley-Dickson  (C-D) algebra. The equation we consider introduces a new interaction term which couples the field to an external source. The term is reminiscent of the interaction of a vector and a spinor field arising in supersymmetric theories. In our case we consider the external field $v$ a constant field valued on the C-D algebra. It could be interpreted as a mean value of the external field. The interaction term, which is zero for the real and complex algebras, depends explicitly on the structure constants of the C-D algebra.  The integrable equation is naturally invariant under the automorphisms of the algebra which preserve the interacting terms. For the octonions the equation in invariant under $SU(3)$. In this sense the new term we introduce is a symmetry breaking term in order to reduce the original global symmetry to a physical relevant one. $SU(3)$ is relevant in the description of fundamental particles.

We notice that the non-associative algebras considered by V. V. Sokolov and S. I. Svinolupov \cite{Sokolov1} do not contain the octonion algebra nor the C-D algebras beyond it. In fact, they consider among others: Lie algebras, Jordan algebras, left-symmetric algebras, L-T algebras. The basic identity they use in their paper is given in equation (0.18) of the above reference. It is straighforward, by considering three elements of the octonionic imaginary basis not belonging to a quaternionic subspace, to verify that (0.18) is not satisfied for the octonion algebra and hence for any Cayley-Dickson algebra beyond it. For example, refering to the Fano plane in section 3, take $e_1,e_3,e_7=e_1e_2$ and use definitions in the refered paper.

We prove in this work that the field equation we introduce here is an
integrable equation in the sense that it has an infinite sequence of conserved quantities.
To do this we will introduce a B\"{a}cklund transformation in the sense of Wahlquist-Estabrook and relate it to a generalized Gardner equation. From it one can directly obtain the infinite sequence of conserved quantities. The product on the transformation is defined in terms of the corresponding algebra. Its existence is a non-trivial result. Besides we obtain the associated Lax pair and obtain from the B\"{a}cklund transformation the one-soliton and two-soliton solutions.

In \cite{Huang,Carillo} an operatorial non-conmuting associative approach to analyze non-linear evolution equations was introduced. In \cite{Huang} multisolitonic solutions for the quaternion KdV equation were found. They have interesting properties beyond the one in the scalar KdV equation, showing the relevance of considering such extension. In \cite{Huang} only two conserved quantities were found. In \cite{Carillo} a Miura transformation was obtained. In the present work we extend the soliton solution to an octonion valued one. We obtain, using explicit properties of the octonion algebra, the one-soliton and two-soliton solutions. They arise directly from the B\"{a}cklund transformation we introduce. Besides we obtain an infinite sequence of conserved quantities for the non-linear evolution equation (2), not only for the octonion valued KdV equation but also for any C-D algebra valued KdV equation. The Miura transformation and a new modified KdV equation valued on any C-D algebra follow from the Gardner equation. It is also an integrable equation.

In connection with physics theories it is important to mention that, in particular, the octonion algebra is directly related to the supersymmetric theories. For example octonion truncations of the Supermembrane theories are interesting models for describing aspects of the unification of the fundamental forces in nature. Recently higher spin constructions have been related to the KdV equation \cite{Troncoso}.  See \cite{Toppan1,Toppan2} for a direct relation of C-D algebras and dimensionally reduced supersymmetric theories.

In Section 2 we define the new equation with values on a general C-D algebra and we
analyze its global symmetries. In Section 3 we consider the particular case of the octonions, we use an explicit representation of the algebra of
the $G_2$ exceptional Lie group of automorphisms of the octonions. In Section 4 we introduce a B\"{a}cklund transformation in the sense of Wahlquist-Estabrook for a KdV equation valued on a C-D algebra. In Section 5 we relate the Wahlquist-Estabrook
construction to a generalized Gardner transformation and Gardner equation, and we give an explicit Lax pair for the equation (2). We also obtain a generalized modified KdV equation. In Section 6 we prove the existence of an infinite sequence of conserved quantities. In Section 7 we obtain, using the B\"{a}cklund transformation, the one and two solitonic solutions for the particular case of the octonion KdV equation. In section 8 we  give the conclusions.

\section{An integrable equation valued on the C-D algebra}
The C-D algebras are constituted by a sequence of algebras, starting with the reals $\mathbb{R}$, obtained inductively with what is called the Cayley-Dickson process. At every stage of this process, a new algebra with twice the dimension of the previous one is formed by considering pairs of elements in the preceding algebra, with multiplication given by $(p,q)(r,s)=(pr-s^*q,sp+qr^*)$ where ${(p,q)}^*=(p^*,-q)$ is the conjugation map and $a^*=a$ for $a\in\mathbb{R}$. The first four algebras generated by this process are, precisely, the normed division algebras: the reals $\mathbb{R}$, the complex numbers $\mathbb{C}$, the quaternions $\mathbb{H}$, and the octonions $\mathbb{O}$. The octonion algebra is a non-commutative, non-associative algebra, and alternative normed division algebra. All further algebras in this process have zero divisors and lack the alternative property; the next algebra in the sequence is known as the sedenions $\mathbb{S}$. Nonetheless, all of the C-D algebras are power associative.

The Cayley-Dickson (C-D) algebra of dimension $2^n$ contains a basis $\{e_0,\,e_1,\ldots,\,e_{2^n-1}\}$ with the following relations:
\[
e_0\,e_0=e_0,
\quad
e_0^*=e_0,
\quad
e_0\,e_i=e_i\,e_0=e_i,
\quad
e_ie_i=-e_0,
\quad
e_i^*=-e_i,
\quad
e_i\,e_j=-e_j\,e_i,
\]
for $i,j\in\{1,\ldots,2^n-1\}$ and $i\neq j$.

We denote by $C_{ijk}$ the structure constants defined by $[e_i,e_j]=\sum_{k=1}^{2^n-1}
C_{ijk}e_k,$ with $C_{ijk}$  being totally antisymmetric on the indices $i,j,k.$

Any element $x$ of the algebra can be expressed into its real and
imaginary parts
\[x=A_0e_0+A_ie_i=a+\vec{A}\] where $a^*=a$ and $
{\vec{A}}^*=-\vec{A}$.

The associator of any three elements $x$, $y$, and $z$ of the algebra is
defined by
\[\left[x,y,z\right]=\left(xy\right)z-x\left(yz\right).\] It is
zero for associative algebras, as $\mathbb{R}$, $\mathbb{C}$, and $\mathbb{H}$. It is skew symmetric for alternative algebras as the
octonion algebra. The C-D algebras beyond the octonions fail to have the alternative property (that is, $(x^2)y\neq x(xy)$ and $(xy)y\neq x(y^2)$, in general). However all of them are power associative
algebras.

In what follows we assume $u=u(x,t)$ a function with domain in $
\mathbb{R}\times\mathbb{R}$ valued on the C-D algebra. If
we denote $e_i,i=1,\ldots,2^n-1$ the imaginary basis of the
C-D algebra, $u$ can be expressed (for each pair of $x$ and $t$ in their domain) as \beq
u(x,t)=b(x,t)+\overrightarrow{B}(x,t)  \eeq where $b(x,t)$ is the
real part and $\overrightarrow{B}=\sum_{i=1}^{2^n-1}B_i(x,t)e_i$ its
imaginary part.

The equation formulated on the C-D algebra is given by
\beq u_t+u_{xxx}+\frac{1}{2}{(u^2)}_x+[v,u]=0,  \eeq where the product in $u^2$ and in $[v,u]$ is the product on the C-D algebra, $v$ is a constant element on the C-D algebra, which can be interpreted as an external field. 

If $v=a+\overrightarrow{A}$ then
\[[v,u]= \left(vu-uv\right)=[a+\overrightarrow{A},b+\overrightarrow{B}]=[\overrightarrow{A},
\overrightarrow{B}]=A_jB_kC_{jki}e_i,\] where $[e_i,e_j]=\sum_{k=1}^{2^n-1}C_{ijk}e_k,$ $C_{ijk}$ are the structure constants characterizing the C-D algebra. For every C-D algebra $C_{ijk}$  are completely antisymmetric on the three indices $i,j,k.$ 

When $\overrightarrow{B}=\overrightarrow{0}$ the equation reduces to the scalar KdV equation. For the real and complex algebras $[v,u]=0$. $[v,u]$ is always pure imaginary.

In terms of $b$ and $\overrightarrow{B}$ the equation can be re-expressed as

\begin{eqnarray}&& b_t+b_{xxx}+bb_x-\sum_{i=1}^{2^n-1}B_iB_{ix}=0,\\ &&
{(B_i)}_t+{(B_i)}_{xxx}+{(bB_i)}_x+\sum_{j,k=1}^{2^n-1}A_jB_kC_{jki}=0.
\end{eqnarray}
In the case $v=0$, the space of solutions of the system (3) and (4) for any $B_i, i=1,\ldots,N$ is contained in the space of solutions of (2) for every Cayley-Dickson algebra with a number of imaginary generators greater than $N$. In fact, if at $t=0$ $B_i=0$ for $i\in I$ then equation (4) implies that $B_i=0$ for all $t$ and $i\in I$. That is, the system (3) and (4) with $B_i, i=1,\ldots,N$ is a subsystem of (2) for any C-D algebra with more imaginary generators than $N$, in a way that all solutions of the subsystem are solutions of the C-D system. The subsystem is not invariant under the automorphisms of the C-D algebra. It is mapped under them into another subsystem whose space of solutions is also contained in the space of solutions of the C-D system (2). The space of solutions of the C-D algebra is manifestly invariant under the automorphisms of it. The converse is also valid: the space of solutions of (2) for a given C-D algebra is contained into the space of solutions of (3) and (4) for a given set $B_i, i=1,\ldots,N$ with $N$ greater than the number of imaginary generators of the C-D algebra.

In the case $v\neq0$ we do not know which is the relation between the system (3) and (4), for any number of $B_i, i=1,\ldots,N$ and antisymmetric constants $C_{ijk}$, and the system associated to a C-D algebra. The reason is that the interactions term $A_jb_kC_{ijk}$ combine in a non-trivial way the different components $B_i$. Certainly, if they are not equivalent, the symmetries of the C-D system are going to be lost, so we lose one of the main properties for describing a physical system.

Equation (2) is invariant under Galileo transformations: 
\begin{eqnarray*}&&\widetilde{x}=x+ct,\\&&\widetilde{t}=t,\\&&\widetilde{u}=u+c, \\&& \widetilde{v}=v,
\end{eqnarray*} where $c$ is a real constant. 

Additionally, equation (2), when $v\neq 0$, is invariant under the subgroup of 
automorphisms of the C-D algebra which preserves $v$. 

If under an automorphism \[ u\rightarrow \phi(u)\] then
\[u_1u_2\rightarrow \phi(u_1u_2)= \phi(u_1)\phi(u_2)\]
and consequently \[
\left[\phi(u)\right]_t+\left[\phi(u)\right]_{xxx}+\frac{1}{2}{\left({\left[\phi(u)\right]}^2\right)}_x=0.\] We explicitly analize the symmetry under the group $G_2$, the automorphisms of the octonions in the next section.

In the case of the octonion algebra when $v=0$, the equation (4) and hence (3), is invariant under the action on $B_i,$ $i=1,\ldots,7$,  by a rotation belonging to $SO(7)$ and hence under $G_2$, a subgroup of $SO(7)$ . In the general case, when $v\ne 0$, the symmetry reduces to the subgroup of $G_2$  which preserves $v$. For example, if $v$ is equal to one of the imaginary generators of the algebra then the subgroup is $SU(3)$  \cite{Arenas}.

\section{Symmetries of the equation in the particular case of the octonions} We examine here in more detail, some relevant symmetries for equation (2) in the case of octonions. 

According to Cartan's classification of simple Lie groups, $G_2$ is the smallest exceptional Lie group. It is the group of automorphisms of the
octonions. The tangent space to a group of automorphisms is an
algebra of derivations. Therefore the Lie algebra $g_2$ of the
Lie group $G_2$ is $Der( \mathbb{O})$. The elements in $Der(\mathbb{O})$ can be expressed as linear combinations of maps $D_{a,b}:\:\mathbb{O}\rightarrow\mathbb{O}$, for $a,b\in\mathbb{O}$, given by
\[D_{a,b}(x)=\frac{1}{2}\left(\left[\left[a,x\right],b\right]+\left[a,\left[b,x\right]\right]+\left[\left[a,b\right],x\right]\right)=
\left[\left[a,b\right],x\right]-3\left[a,b,x\right]\]
where $\left[a,b\right]=ab-ba$ is the conmutator and the bracket with three
entries is the associator
$\left[a,b,x\right]=\left(ab\right)x-a\left(bx\right)$. The associator for the octonionic algebra is
totally antisymmetric.

The map satisfies the Leibniz rule
\[D_{a,b}(xy)=D_{a,b}(x)y+xD_{a,b}(y)\] and the generalized Jacobi
identity
\[D_{a,b}\left(D_{c,d}\right)=D_{D_{a,b}(c),d}+D_{c,D_{a,b}(d)}+D_{c,d}(D_{a,b}).\]
We denote the pure imaginary basis of the octonionic algebra by
$e_i,i=1,\ldots,7$ and its multiplication rule is represented by the Fano plane in Figure 1.

\begin{figure}
\begin{center}
    \includegraphics[scale=0.35]{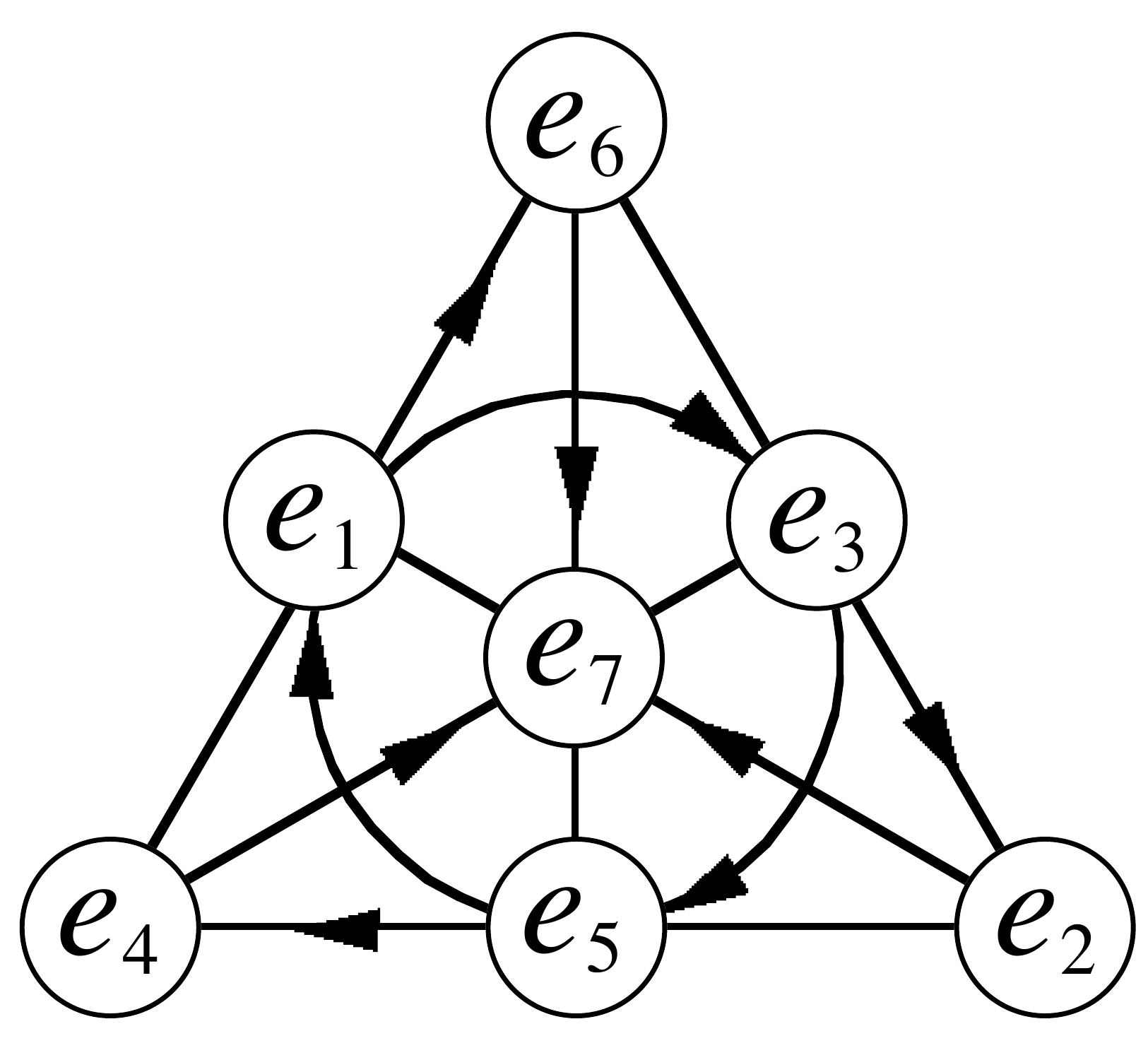}
\end{center}
\caption{Fano plane representing the multiplication rule for the imaginary part of the basis for the octonions, with chosen labels.}
\end{figure}

For each index $p\in\{1,\ldots,7\}$ consider the three pairs of
indices $(i,j),(r,s)$ and $(u,v)$ such that
$e_p=e_ie_j=e_re_s=e_ue_v$ then \beq
D_{e_i,e_j}+D_{e_r,e_s}+D_{e_u,e_v}=0 \eeq is identically
satisfied.

A basis for $g_2$ can be given by fourteen of these maps grouped in
seven pairs, each of them constituted by $D_{e_i,e_j}$ and
$D_{e_r,e_s}$ where $e_ie_j=e_re_s=e_p$ for each
$p\in\{1,\ldots,7\}$.

Let us now consider the transformation of the octonionic KdV
equation under an infinitesimal $G_2$ transformation \beq
u\rightarrow u+\sum_{i,j}\lambda^{ij}D_{e_i,e_j}(u)\eeq where
$\lambda^{ij}=-\lambda^{ji}$ are the infinitesimal parameters of
the transformation and the summation is only over the fourteen elements
of the basis of $g_2$. The parameters $\lambda^{ij}$ are
independent of $(x,t)$ since the $G_2$ symmetry is a global one.

The transformation of (2) under (6) is
\[\phi\equiv{\left(\delta u\right)}_t+{\left(\delta u\right)}_{xxx}+\frac{1}{2}{\left(\delta u\cdot u+u\cdot \delta
u\right)}_x+[\delta v,u]+[v,\delta u]\]

where $\delta
u=\Sigma_{i,j}\lambda^{ij}D_{e_i,e_j}(u).$

Using now the Leibnitz rule for the derivation map we obtain
\[\phi=\lambda^{ij}D_{e_i,e_j}\left(u_t+u_{xxx}+\frac{1}{2}{(u^2)}_x+[v,u]\right)=0\]
which shows that Equation (2) is invariant under the
transformation given in (6) if $v=0$ and under the subgroup $SU(3)$ if $v$ is equal to one of the imaginary generators of the octonion algebra.

\section{B\"{a}cklund transformation for the Cayley-Dickson 
equation}
The Wahlquist-Estabrook (WE) transformation \cite{Wahlquist} for the
real KdV equation can be straightforwardly generalized for the
complex KdV one. We prove, in what follows, its extension
for the C-D algebras. It is a non-trivial result due to the new interacting term and to the fact that the field is valued in general on a non-associative, non-alternative algebra. 

We define $w$ through $u=w_x$. From (2) we obtain
\[Q(w)\equiv w_t+w_{xxx}+\frac{1}{2}{(w_x)}^2+[v,w]=C(t)\]
where $C(t)$ is a function of $t$ only. As in the case of the
real KdV equation we can redefine
\[\widetilde{w}=w-\int_{-\infty}^tC(\tau)d\tau.\] We then obtain
$Q( \widetilde{w})=0$. In what follows we assume that this
redefinition has been performed and consider $Q(w)=0$ in order to
simplify the notation.

The B\"{a}cklund transformation for the C-D equation has the
expression
\begin{eqnarray}&&
w_x+w_x^\prime=\eta-\frac{1}{12}{\left(w-w^\prime\right)}^2 \\ &&
w_t+w_t^\prime=\frac{1}{12}{\left[{\left(w-w^\prime\right)}^2
	\right]}_{xx}-\frac{1}{2}{w_x}^2-\frac{1}{2}{w_x^\prime}^2-[v,w+w^\prime]
\end{eqnarray} where $w=w(x,t)$ and $w^\prime=w^\prime(x,t)$ are
valued on the C-D algebra.

We now prove the following proposition.

\begin{proposicion}If $w$ and $w^\prime$ are solutions of (7) and
(8), satisfying $\mathbb{R}e(w-w^\prime)\neq0$, then $u=w_x$ and
$u^\prime=w_x^\prime$ are solutions of the C-D equation.
\end{proposicion}
\begin{demosprop}We consider $(7)_{xx}+(8),$  and obtain
\beq Q(w)+Q(w^\prime)=0.\eeq

We now consider the integrability condition for (7) and (8):
\[(7)_t-(8)_x=0.\] 

We get,
\beq \begin{array}{lll}-\frac{1}{12}
\left[{\left(w-w^\prime\right)}_t\left(w-w^\prime\right)+
\left(w-w^\prime\right)
{\left(w-w^\prime\right)}_t\right]-\frac{1}{12}
\left[{\left(w-w^\prime\right)}^2\right]_{xxx}+
\\  \\
+\frac{1}{2}\left[w_{xx}w_x+w_xw_{xx}\right]+\frac{1}{2}\left[w_{xx}^\prime
w_x^\prime+w_x^\prime w_{xx}^\prime\right]+[v,w_x+w_x^\prime]=0.\end{array} \eeq

Also \beq
\begin{array}{lll}-\frac{1}{12}\left[{\left(w-w^\prime\right)}^2\right]_{xxx}
	=-\frac{1}{12}{\left(w-w^\prime\right)}_{xxx}\left(w-w^\prime\right)-
\frac{1}{4}{\left(w-w^\prime\right)}_{xx}{\left(w-w^\prime\right)}_x-
\\ \\
-\frac{1}{4}{\left(w-w^\prime\right)}_x{\left(w-w^\prime\right)}_{xx}-\frac{1}{12}\left(w-w^\prime\right){\left(w-w^\prime\right)}_{xxx}.
\end{array}\eeq
The fourth and fifth  terms of (10) combine with the second and
third terms of the right hand side member of equation (11) to give
\beq
\begin{array}{lll}\frac{1}{2}\left[w_{xx}w_x+w_xw_{xx}\right]+\frac{1}{2}\left[w_{xx}^\prime w_x^\prime+w_x^\prime w_{xx}^\prime\right]
-\frac{1}{4}{\left(w-w^\prime\right)}_{xx}{\left(w-w^\prime\right)}_x-
\\ \\
-\frac{1}{4}{\left(w-w^\prime\right)}_x{\left(w-w^\prime\right)}_{xx}=\frac{1}{4}\left(w_{xx}+w^\prime_{xx}\right)\left(w_x+w^\prime_x\right)+
\frac{1}{4}\left(w_x+w^\prime_x\right)\left(w_{xx}+w^\prime_{xx}\right).
\end{array} \eeq
We may now use (7) to obtain
\beq \begin{array}{lll} \frac{1}{4}\left(w_{xx}+w^\prime_{xx}\right)\left(w_x+w^\prime_x\right)+
\frac{1}{4}\left(w_x+w^\prime_x\right)\left(w_{xx}+w^\prime_{xx}\right)=\\=-\frac{1}{48}\left[{\left(w-w^\prime\right)}_x
\left(w-w^\prime\right)+\left(w-w^\prime\right){\left(w-w^\prime\right)}_x\right]\left(w_x+w^\prime_x\right)-\\
-\frac{1}{48}\left(w_x+w^\prime_x\right)\left[{\left(w-w^\prime\right)}_x
\left(w-w^\prime\right)+\left(w-w^\prime\right){\left(w-w^\prime\right)}_x\right].
\end{array}\eeq
Furthermore, using the definition of the associator,
\begin{eqnarray*}&&\left(w_x+w^\prime_x\right)\left[{\left(w-w^\prime\right)}_x
\left(w-w^\prime\right)\right]=\left[\left(w_x+w^\prime_x\right)\left(w_x-w^\prime_x\right)\right]\left(w-w^\prime\right)-
\left[w_x+w^\prime_x,w_x-w^\prime_x,w-w^\prime\right], \\ &&
\left[{\left(w-w^\prime\right)}
\left(w-w^\prime\right)_x\right]\left(w_x+w^\prime_x\right)=\left(w-w^\prime\right)\left[\left(w_x-w^\prime_x\right)
\left(w_x+w^\prime_x\right)\right]+
\left[w-w^\prime,w_x-w^\prime_x,w_x+w_x^\prime\right].
\end{eqnarray*}

Summation of the right hand members of these two equations yields
\begin{eqnarray*}&& \left[{(w_x)}^2-{(w_x^\prime)}^2\right]\left(w-w^\prime\right)+\left(w-w^\prime\right)\left[{(w_x)}^2-{(w_x^\prime)}^2\right]
+\left(w_x^\prime w_x-w_x w_x^\prime\right)\left(w-w^\prime\right)+\\&&+\left(w-w^\prime\right)
\left(w_x w_x^\prime-w_x^\prime w_x\right)+\left[w-w^\prime,w_x-w^\prime_x,w_x+w_x^\prime\right]-
\left[w_x+w_x^\prime,w_x-w^\prime_x,w-w^\prime\right].
\end{eqnarray*}
In the same way
\begin{eqnarray*}&&\left[\left(w_x-w_x^\prime\right)\left(w-w^\prime\right)\right]\left(w_x+w^\prime_x\right)=
\left(w_x-w^\prime_x\right)\left[\left(w-w^\prime\right)\left(w_x+w_x^\prime\right)\right]+
\left[w_x-w_x^\prime,w-w^\prime,w_x+w_x^\prime\right],
\\&&\left(w_x+w^\prime_x\right)
\left[\left(w-w^\prime\right)\left(w_x-w_x^\prime\right)\right]=\left[\left(w_x+w_x^\prime\right)
\left(w-w^\prime\right)\right]\left(w_x-w^\prime_x\right)-\left[w_x+w_x^\prime,w-w^\prime,w_x-w_x^\prime\right].
\end{eqnarray*}
Using (7) and the power associative property of the
Cayley-Dickson algebras we get
\begin{eqnarray*}&&\left(w_x-w^\prime_x\right)\left[\left(w-w^\prime\right)
\left(w_x+w^\prime_x\right)\right]=\left(w_x-w^\prime_x\right)\left[\left(w_x+w_x^\prime\right)
\left(w-w^\prime\right)\right]=\left[\left(w_x-w_x^\prime\right)
\left(w_x+w^\prime_x\right)\right]\cdot \\&&\cdot \left(w-w^\prime\right)-
\left[w_x-w_x^\prime,w_x+w^\prime_x,w-w^\prime\right]=\left({\left(w_x\right)}^2-{\left(w_x^\prime\right)}^2\right)
\left(w-w^\prime\right)+\\&&+\left(w_xw_x^\prime-w_x^\prime w_x\right)\left(w-w^\prime\right)-
\left[w_x-w_x^\prime,w_x+w^\prime_x,w-w^\prime\right],\end{eqnarray*}
\begin{eqnarray*}&&\left[\left(w_x+w_x^\prime\right)
\left(w-w^\prime\right)\right]\left(w_x-w^\prime_x\right)=\left[\left(w-w^\prime\right)
\left(w_x+w_x^\prime\right)\right]\left(w_x-w^\prime_x\right)=\\&&
\left(w-w^\prime\right)\left({\left(w_x\right)}^2-{\left(w_x^\prime\right)}^2\right)+
\left(w-w^\prime\right)\left(w_x^\prime w_x-w_x
w_x^\prime\right)+\left[w-w^\prime,w_x+w^\prime_x,w_x-w_x^\prime\right].\end{eqnarray*}
Summation of the right hand members of these two equations yields
\begin{eqnarray*}&&\left[{\left(w_x\right)}^2-{\left(w_x^\prime\right)}^2\right]\left(w-w^\prime\right)+
\left(w-w^\prime\right)\left[{\left(w_x\right)}^2-{\left(w_x^\prime\right)}^2\right]+\left(w_x
w_x^\prime-w_x^\prime
w_x\right)\left(w-w^\prime\right)+\\&&+\left(w-w^\prime\right)\left(w_x^\prime
w_x-w_x w_x^\prime\right)-
\left[w_x-w_x^\prime,w_x+w^\prime_x,w-w^\prime\right]+\left[w-w^\prime,w_x+w^\prime_x,w_x-w_x^\prime\right].
\end{eqnarray*}

We then have from (13) \beq \begin{array}{llll}
\frac{1}{4}\left(w_{xx}+w_{xx}^\prime\right)\left(w_x+w_x^\prime\right)+
\frac{1}{4}\left(w_{x}+w_{x}^\prime\right)\left(w_{xx}+w_{xx}^\prime\right)=
-\frac{1}{24}\left({\left(w_x\right)}^2-{\left(w_x^\prime\right)}^2\right)\left(w-w^\prime\right)-
\\
-\frac{1}{24}\left(w-w^\prime\right)\left({\left(w_x\right)}^2-{\left(w_x^\prime\right)}^2\right)-
\frac{1}{48}\{\left[w-w^\prime,w_x-w^\prime_x,w_x+w_x^\prime\right]-\\-
\left[w_x+w_x^\prime,w_x-w^\prime_x,w-w^\prime\right]-\left[w_x-w_x^\prime,w_x+w^\prime_x,w-w^\prime\right]
+\left[w-w^\prime,w_x+w^\prime_x,w_x-w_x^\prime\right]+\\+\left[w_x-w_x^\prime,w-w^\prime,w_x+w_x^\prime\right]-
\left[w_x+w_x^\prime,w-w^\prime,w_x-w_x^\prime\right]\}.
\end{array}\eeq

We may now evaluate explicitly the associators, to do so we
replace $w_x+w_x^\prime$ by its expression given from (7), for example
\[\left[w_x-w_x^\prime,w_x+w^\prime_x,w-w^\prime\right]=\left[w_x-w_x^\prime,\eta,w-w^\prime\right]-
\frac{1}{12}\left[w_x-w_x^\prime,{(w-w^\prime)}^2,w-w^\prime\right].\]
Since $\eta$ is real, the first term of the right hand side member
is zero. The second term involves $w-w^\prime=a+\vec{v}$, where
$a$ is its real part and $ \vec{v}$ its imaginary one. We have,
using the properties of the basis for any C-D algebra,
\begin{eqnarray*}&&{\left(w-w^\prime\right)}^2=a^2+2a\vec{v}-{\|\vec{v}\|}^2, \\&&
\left[w_x-w_x^\prime,a^2+2a\vec{v}-{\|\vec{v}\|}^2,a+\vec{v}\right]=
\left[w_x-w_x^\prime,2a\vec{v},\vec{v}\right]=2a\left[w_x-w_x^\prime,\vec{v},\vec{v}\right],\end{eqnarray*}
which is zero for any alternative algebra like the octonions but
it is not zero for a generic C-D algebra.

However, we notice that using this result and the corresponding
ones for the other associators in (14), the associators cancel by
pairs: the first with the second, the third with the fifth and
the fourth with the sixth.

For the latter term in (10) we have, using (7) and denoting $w-w^\prime=d+\vec{D}$ and $v=a+\vec{A},$
\[[v,w_x+w_x^\prime]=-\frac{1}{12}\left[v,{(w-w^\prime)}^2\right]=-\frac{1}{6}
d[\vec{A},\vec{D}]\] also
\[-\frac{1}{12}(w-w^\prime)\left( \left[v,w-w^\prime\right]\right)-\frac{1}{12}
\left( \left[v,w-w^\prime\right]\right)(w-w^\prime)=\]
\[=-\frac{1}{12}(d+\vec{D})\left( [\vec{A},\vec{D}]\right)-\frac{1}{12}\left( [\vec{A},\vec{D}]\right)(d+\vec{D})=-\frac{1}{6}d[\vec{A},\vec{D}].\]

We then obtain \beq
\left[Q(w)-Q(w^\prime)\right]\left(w-w^\prime\right)+\left(w-w^\prime\right)\left[Q(w)-Q(w^\prime)\right]=0.\eeq
If the real part of $w-w^\prime$ is different from zero:
$\mathbb{R}e\left(w-w^\prime\right)\neq0$, the above equation
implies \beq Q(w)-Q(w^\prime)=0.  \eeq In fact, if we denote
\begin{eqnarray*}&&w-w^\prime=a+\vec{v},
\\&&Q(w)-Q(w^\prime)=c+\vec{w},
\end{eqnarray*} the decomposition into its real and imaginary parts, then $(a+\vec{v})(c+\vec{w})+(c+\vec{w})(a+\vec{v})=0$
imply that its real part is zero and so is its imaginary part.

That is,
\begin{eqnarray*}&&
2ac+\vec{v}\vec{w}+\vec{w}\vec{v}=0,\\&&2a\vec{w}+2c\vec{v}=0,\end{eqnarray*}
where due to the properties of the basis of the C-D algebra
$\vec{v}\vec{w}+\vec{w}\vec{v}$ is real.

If $a\neq0$, we then get
\begin{eqnarray*}&&\vec{w}=-\frac{c}{a}\vec{v}, \\&&2ac+\vec{v}\vec{w}+\vec{w}\vec{v}=\frac{2c}{a}\left(a^2-\vec{v}
\vec{v}\right)=0
\end{eqnarray*} but $a^2-\vec{v}
\vec{v}=a^2+{\|\vec{v}\|}^2\neq0$, hence we must have $c=0$ and
$\vec{w}=0$, that is, equation (16).

Finally, from (9) and (16) we get
\[Q(w)=Q( w^\prime)=0\] and hence $u=w_x$ and $u^\prime=w_x^\prime$ are both solutions of the C-D KdV equation.
\end{demosprop}
\underline{Remark 1}: The converse of Proposition 1 is not valid. Given $u$ and $u^\prime$ solutions of (2) then
$w,w^\prime$ defined through $u=w_x,u^\prime=w_x^\prime$ respectively, do not generally satisfy equations (7) and (8).
\section{The B\"{a}cklund transformation, the generalized Gardner equation and the Lax pair}
We will assume, as in Proposition 1, $
\mathbb{R}e(w-w^\prime)\neq0$. In the previous section we showed
that equations (7) and (8) imply
\[Q(w)-Q(w^\prime)=0.\]
We have
\[Q(w)-Q(w^\prime)=\left(w-w^\prime\right)_t+\left(w-w^\prime\right)_{xxx}+
\frac{1}{2}{\left(w_x\right)}^2
-\frac{1}{2}{\left(w_x^\prime\right)}^2+[v,w-w^\prime]\] where
\[{\left(w_x\right)}^2
-{\left(w_x^\prime\right)}^2=\frac{1}{2}\left(w_x+w_x^\prime\right)\left(w_x-w_x^\prime\right)+
\frac{1}{2}\left(w_x-w_x^\prime\right)\left(w_x+w_x^\prime\right).\]
We may now use (7) to obtain \beq \begin{array}{ll}
\left(w-w^\prime\right)_t+\left(w-w^\prime\right)_{xxx}
+\frac{1}{4}\left[\eta-\frac{1}{12}
{\left(w-w^\prime\right)}^2\right]
\left(w_x-w_x^\prime\right)+ \\+\frac{1}{4}\left(w_x-w_x^\prime\right)\left[\eta-\frac{1}{12}
{\left(w-w^\prime\right)}^2\right]+[v,w-w^\prime]=0.\end{array}\eeq We introduce the field
$r(x,t)$ through the relation
$w-w^\prime=2\epsilon\left(r-\frac{3}{\epsilon^2}\right)$ where
$\epsilon\neq0$ is a real parameter. We get
\[\eta-\frac{1}{12}
{\left(w-w^\prime\right)}^2=-\frac{1}{3}\epsilon^2\left(r^2-\frac{6r}{\epsilon^2}\right)\] provided $\eta=\frac{3}{\epsilon^2}.$

We finally obtain from (17) \beq
r_t+r_{xxx}+\frac{1}{2}\left(rr_x+r_xr\right)-\frac{1}{12}\left((r^2)r_x+r_x(r^2)\right)\epsilon^2+[v,r]=0.
\eeq

This equation, where $r(x,t)$ is valued on the C-D algebra and the product is the one of the algebra, is the
generalized Gardner equation. It allows us to obtain an infinite
sequence of conserved quantities, as we will show in the next
section.
\begin{proposicion}If $w$ and $w^\prime$ are solutions of the B\"{a}cklund equations (7) and (8)
and $ \mathbb{R}e(w-w^\prime)\neq0$, then
$r=\frac{3}{\epsilon^2}+\frac{1}{2\epsilon}\left(w-w^\prime\right)$
satisfy the generalized Gardner equation (18) and
\begin{eqnarray}u=w_x=r+\epsilon r_x-\frac{1}{6}\epsilon^2r^2 \\u^\prime=
w_x^\prime=r-\epsilon r_x-\frac{1}{6}\epsilon^2r^2 \end{eqnarray}
are solutions of the C-D KdV equation.

If $r(x,t)$ is a solution of the generalized Gardner equation
(18), then $u$ and $u^\prime$ given by (19) and (20) are solutions
of the C-D equation and $w,w^\prime$ are solutions, by
choosing the integration constants in a way that
$r=\frac{3}{\epsilon^2}+\frac{1}{2\epsilon}(w-w^\prime)$, of the
B\"{a}cklund equations (7) and (8).
\end{proposicion}
\underline{Remark 2} We notice that $u^\prime$ is obtained from $u$ by changing $\epsilon\rightarrow-\epsilon$.
\begin{demosprop}From the previous argument in this section $r(x,t)$ is a solution of the Gardner equation. In addition, from (7)
\beq w_x+w_x^\prime=2r-\frac{1}{3}\epsilon^2r^2,   \eeq  and from
the definition of $r$ in terms of $w-w^\prime$ \beq
w_x-w_x^\prime=2\epsilon r_x.\eeq From these two equations we
obtain (19) and (20). Proposition 1 ensures that $u,u^\prime$ are
solutions of (2).

If $r(x,t)$ is a solution of the Gardner equation then defining
$w$ and $w^\prime$ from (19) and (20) we conclude that (21) and
(22) are satisfied. By fixing the integration constant obtained
from (22) in a way that
\[r=\frac{3}{\epsilon^2}+\frac{1}{2\epsilon}(w-w^\prime),\] (17) is then satisfied. Since this is the integrability
condition of (7) and (8) we obtain that $w$ and $w^\prime$ are
solutions of (7) and (8) (we notice that (7) arises directly from
(21)). We now apply Proposition 1 to show that $u$ and $u^\prime$
given by (19) and (20) are solutions of (2).
\end{demosprop}

Lax introduced an approach to construct a large class of nonlinear evolution equations with multisolitonic solutions. Besides, the B\"{a}cklund transformation is a natural approach to obtain multisolitonic solutions starting from trivial ones.

In \cite{Chen} a general way of obtaining the B\"{a}cklund transformation from the Lax equations was found. Following these ideas one can obtain the Lax pair for the nonlinear equation (2) valued on a C-D algebra from the B\"{a}cklund transformation we have considered. It is given by
\begin{eqnarray*}
	&&\mathcal{L}_t=\left[\mathcal{L},\mathcal{P}\right], \\
	&&\mathcal{L}=-\partial^2_x-\frac{1}{6}u,\\ && \mathcal{P}
	=4\partial^3_x+u\partial_x+\frac{1}{2}\partial_xu+6v,\end{eqnarray*} defined for $u(x,t)$
and $v$ valued on a C-D algebra. The eigenvalued problem for the operator 
$\mathcal{L}$, however, is well defined on a Hilbert space constructed for division algebras only. Hence, this Lax pair can in principle be formulated up to the octonion algebra but not beyond it.

From (18) we obtain directly the modified KdV equation. First we multiply by $\epsilon$, redefine $\hat{r}=\epsilon r$ and take the limit $\epsilon \rightarrow \infty.$

We obtain,
\[\hat{r}+{\hat{r}}_{xxx}-\frac{1}{12}\left({\left(\hat{r}^2\right)\hat{r}_x}+\hat{r}_x
\left(\hat{r}^2\right)\right)+\left[v,\hat{r}\right]=0,\] the modified KdV equation defined on a C-D algebra, with $\hat{r}$ valued on it. The product is understood as the product in the C-D algebra.

\section{An infinite sequence of conserved quantities}We first show that $\int_{-\infty}^{+\infty}\mathbb{R}e
\left[r\left(x,t\right)\right]dx$ is a conserved quantity of the generalized Gardner equation (18).
\begin{proposicion}Let $r(x,t)$ be a solution of the Gardner equation (18) and assume $r(x,t)\in\mathcal{L}( \mathbb{R})$,
the Schwartz space of functions on $ \mathbb{R}$, then $\int_{-\infty}^{+\infty}\mathbb{R}e
\left[r\left(x,t\right)\right]dx$ is a conserved quantity.  \end{proposicion}
\begin{demosprop}Taking the real part of equation (18), we obtain
\[\left[\mathbb{R}e
\left(r\right)\right]_t+\left[\mathbb{R}e
\left(r\right)\right]_{xxx}+\frac{1}{2}\left[\mathbb{R}e
\left(r^2\right)\right]_{x}-\frac{1}{12}\epsilon^2\mathbb{R}e
\left(r^2r_x+r_xr^2\right)=0.\]
We now show that $\mathbb{R}e
\left(r^2r_x+r_xr^2\right)$ is a total derivative. In fact, we consider $r=a+\vec{A}$ where $a$
is the real part of $r$ and $\vec{A}$ its imaginary part. We get
\begin{eqnarray*}&& r^2=a^2+\vec{A}\vec{A}+2a\vec{A}\hspace{2mm},\hspace{2mm}\vec{A}\vec{A}=-{\|\vec{A}\|}^2,
\\&&r^2r_x+r_xr^2=\left(a^2-{\|\vec{A}\|}^2+2a\vec{A}\right)\left(a_x+{\vec{A}}_x\right)+
\left(a_x+{\vec{A}}_x\right)\left(a^2-{\|\vec{A}\|}^2+2a\vec{A}\right)=\\&&=2a^2a_x-2{\|\vec{A}\|}^2
a_x+2a\left(\vec{A}{\vec{A}}_x+{\vec{A}}_x\vec{A}\right)+2a^2{\vec{A}}_x+2aa_x\vec{A}-2{\|\vec{A}\|}^2
{\vec{A}}_x, \\&& \mathbb{R}e
\left(r^2r_x+r_xr^2\right)=\frac{2}{3}{\left(a^3\right)}_x-2
{\|\vec{A}\|}^2a_x-2a{\left({\|\vec{A}\|}^2\right)}_x={\left(\frac{2}{3}a^3-2a{\|\vec{A}\|}^2\right)}_x.\end{eqnarray*}
Then \begin{eqnarray*}&&\frac{d}{dt}\int_{-\infty}^{+\infty}\mathbb{R}e\left[r(x,t)\right]dx=
\\&=&\int_{-\infty}^{+\infty}\left[-{\left(\mathbb{R}e(r)\right)}_{xxx}-\frac{1}{2}
{\left(\mathbb{R}e(r^2)\right)}_{x}+\frac{1}{12}\epsilon^2{\left(\frac{2}{3}
{\left(\mathbb{R}e(r)\right)}^3-2\mathbb{R}e(r){\|\mathbb{I}m(r)\|}^2\right)}_x\right]dx=0.\end{eqnarray*}
\end{demosprop}
We are now able to construct, following a very well known
approach \cite{Miura1,Kupershmidt,Mathieu1,Adrian1,Adrian2,Adrian3}, an infinite sequence of conserved
quantities. Assuming a formal expansion of $r$ in powers of
$\epsilon$, we can invert (19). We obtain
\begin{eqnarray*}&& r=u-u_x\epsilon+\left(u_{xx}+\frac{1}{6}u^2\right)\epsilon^2-
{\left(
u_{xx}+\frac{1}{3}u^2\right)}_x\epsilon^3+\\&&+{\left(
u_{xx}+\frac{1}{3}u^2\right)}_{xx}\epsilon^4+\frac{1}{6}\left(uu_{xx}+\frac{1}{3}u^3+u_{xx}u+
{\left(u_x\right)}^2\right)\epsilon^4+\cdots\end{eqnarray*}

Using now Proposition 3 we get an infinite sequence of conserved
quantities.

The first few of them are
\begin{eqnarray*}&& H_1=\int_{-\infty}^{+\infty}\mathbb{R}e(u)dx,\\&& H_2=\int_{-\infty}^{+\infty}
\left({\left(\mathbb{R}e(u)\right)}^2-{\|\mathbb{I}m(u)\|}^2\right)dx,\\&&H_3=
\int_{-\infty}^{+\infty}\left(\frac{1}{3}{\left(\mathbb{R}e(u)\right)}^3-{\left(\mathbb{R}e(u_x)\right)}^2
+{\|\mathbb{I}m(u_x)\|}^2-\mathbb{R}e(u){\|\mathbb{I}m(u)\|}^2\right)dx.
\end{eqnarray*}

Also, if $v=0$, we have that $\int_{-\infty}^{+\infty}\mathbb{I}m(u)dx$ is a conserved quantity.

In addition to the above conserved quantities, valid for any solution $r(x,t)$ of the Gardner equation, it is also valid the following property for particular solutions of the equation.

Assuming $v=0$ and that there exist solutions of the Gardner equation for which ${\|\mathbb{I}m(r)\|}^2$ is constant. That is, if $r=a+\vec{A}$, suppose there are solutions such that $\vec{A}{\vec{A}}_x+{\vec{A}}_x\vec{A}=0.$  Then 
$\int_{-\infty}^{+\infty}\mathbb{I}m(r)dx$ is conserved. In fact,
for these particular solutions we have
\[\left(r^2\right)r_x+r_x\left(r^2\right)={\left(r^3\right)}_x-
{\left(2{\|\mathbb{I}m(r)\|}^2\vec{A}\right)}_x,\] where $r^3=a^3-3a{\|\mathbb{I}m(r)\|}^2+3a^2\vec{A}-{\|\mathbb{I}m(r)\|}^2\vec{A}$
and ${\|\mathbb{I}m(r)\|}^2=-\vec{A}\vec{A}$.

\section{Soliton solutions from the B\"{a}cklund transformation}In this section we obtain the one-soliton and two-soliton solutions of the octonion KdV equation with and without interaction term $\left[v,\cdot\right]$. As we emphasized earlier this is a symmetry breaking term in order to reduce the symmetry group, preserving the KdV equation, to $SU(3)$ (the group related to the physics of quarks). 

As it is well known, the inverse of an octonion $\alpha$ ($\alpha\in \mathbb{O}$), is
$\frac{1}{\alpha}\equiv \frac{\bar{\alpha}}{{\mid \alpha \mid}^2}$, where  $\alpha=\alpha_0+\vec{\alpha}$ and $\bar{\alpha}=\alpha_0-\vec{\alpha}$, $\vec{\alpha}$ is its pure imaginary part which is expressed in terms of the basis elements $e_1,\ldots, e_7$ and $\alpha_0$ is its real part.

If $F(x,t)$ is an octonion valued function with values different from zero then 
\[\partial_x\left(\frac{1}{F}\right)=-\frac{1}{2}(\partial_xF)\frac{1}{F^2}-\frac{1}{2}
\left(\frac{1}{F^2}\right)(\partial_xF),\] the associator of the three factors in this expression is zero.

We consider $f=e^{-\lambda x+\lambda^3t}$ a wave function as usual in the construction of solitary waves. We then notice that $w_1=6\lambda\left(\frac{1}{\alpha+f}\right)
\left(\alpha-f\right),$ $\alpha\in \mathbb{O}\backslash \mathbb{R}_- $, is a regular solution of the B\"{a}cklund equation (7). In fact, the left hand member of (7) is 
\beq \begin{array}{ll}w_{1x}=6\lambda\left[{\left(\frac{1}{\alpha+f}\right)}_x(\alpha-f)+
	{\left(\frac{1}{\alpha+f}\right)}{(\alpha-f)}_x
	\right] =6\lambda\left(\frac{-f_x}{{(\alpha+f)}^2}\right)(\alpha-f)+\left(\frac{1}{\alpha+f}\right)\cdot  (-f_x)\\ =-6\lambda f_x\left(\frac{1}{{(\alpha+f)}^2}\right)
	(\alpha-f+\alpha+f)=\frac{12{\lambda}^2\alpha f}{{(\alpha+f)}^2},\end{array}\eeq
where no ambiguities arise because only $\alpha$ and $\bar{\alpha}$ are involved in the expression, hence any associator is zero.

On the other side, the right hand member of (7) is
\[\eta-\frac{1}{12}36{\lambda}^2{\left[\left(\frac{1}{(\alpha+f)}(\alpha-f)\right)\right]}^2=
\eta-3{\lambda}^2\left(\frac{1}{{\left(\alpha+f\right)}^2}\right){\left(\alpha-f\right)}^2,\]
since $\frac{1}{\alpha+f}$ and $(\alpha-f)$ conmute and the associator of any three octonions in the expression is zero. 

If $\eta=3{\lambda}^2$ the above  expression reduces to 
\[3{\lambda}^2{\left(\frac{1}{\alpha+f}\right)}^2{\left(\alpha+f\right)}^2-
3{\lambda}^2{\left(\frac{1}{{\left(\alpha+f\right)}^2}\right)}{\left(\alpha-f\right)}^2=
\frac{12{\lambda}^2\alpha f}{{(\alpha+f)}^2},\] which is the same as the left hand member of (7).

When $\alpha$ is real and positive the expression $u_1=w_{1x}$ is the one-soliton solution for the scalar KdV equation. The presence of an octonion in the expression introduces non-trivial properties on the multisolitonic solution. In the particular case in which $\alpha$ is a quaternion this solution was found in \cite{Huang}. In that work it arises from a non-conmutative operatorial approach which is valid for the quaternion algebra but not valid for the non-associative octonion algebra. In our case, besides its extension to a more general algebra, it arises as a first step analysis of the B\"{a}cklund equations. The multisoliton solutions in \cite{Huang} have interesting and non-trivial properties, not present in the scalar case.

It is straighforward to show that $w_2=-6\lambda\left(\frac{1}{f-\alpha}\right)(f+\alpha)$ is also a solution of (7) (change $\alpha$ to $-\alpha$). If $\alpha\in \mathbb{O}\backslash 
\mathbb{R}$ then $w_1$ and $w_2$ are regular solutions of (7). In distinction when $\alpha$ is real and different from zero then one of them, $w_1$ or $w_2$, is singular at some value of $x$.

From (7) it follows that $u_1=w_{1x}$ and $u_2=w_{2x}$ are solutions of 
\beq u_{xx}+\frac{1}{2}u^2-\lambda^2u=0\eeq and of the KdV equation
\beq u_t+u_{xxx}+\frac{1}{2}{\left(u^2\right)}_x=0\eeq
Also if $\alpha=\alpha_0+\vec{\alpha}$ satisfies $\vec{\alpha}=\gamma \vec{v}$, $\gamma$ real, then $u_1$ and $u_2$ are solutions of equation (2). In that case $\left[v,u\right]=0$ and $v$ reduces the symmetry group of the space of solutions to $SU(3)$.

The two-soliton solution is obtained from (7) and the permutability property of the B\"{a}cklund transformation,

\[w=12\frac{\left(\eta_2-\eta_1\right)}{w_2-w_1},\] 
$\eta_1$ and $\eta_2$ are the parameters associated to $u_1$ and $u_2$ respectively.

It can be verified, after several calculations where careful treatment of the associator of some expressions has to be considered, that $u=w_x$ satisfies (25).

It generalizes the two soliton solution of the scalar and quaternion \cite{Huang} solutions. Moreover, it can be shown that for any octonions $\alpha$ and $\beta$, $u=w_x$ with
\[w=12\left(\eta_\alpha-\eta_\beta\right)\left(\frac{1}{w_\alpha-w_\beta}\right),\]
satisfies (25), where
\[w_\alpha=6\lambda_\alpha\left(\frac{1}{\alpha+f_\alpha}\right)\left(f-\alpha\right),\]
\[f_\alpha=\exp\left(-\lambda_\alpha x+\lambda^3_\alpha t\right),\hspace{5mm}
\eta_\alpha=3\lambda^3_\alpha\]  and the analogous expression for $w_\beta$ in terms of the parameter $\lambda_\beta$ and the octonion $\beta$.

When $\vec{\alpha}=\gamma_1\vec{v}, \vec{\beta}=\gamma_2\vec{v}$ for any real $\gamma_1,\gamma_2$ and $\alpha_0,\beta_0, u$ is also a solution of equation (2). In this case we have a six parameter space of solutions: the real parts of the octonions $\alpha_0,\beta_0,$ the scaling factors $\gamma_1,\gamma_2$ and the soliton parameters $\lambda_\alpha$ and $\lambda_\beta$.

 \section{Conclusions}We introduced a new integrable equation valued on a general Cayley-Dickson algebra. The non-linear equation reduces to the complex KdV equation when the algebra reduces to the complex one. One of the interacting terms, the self interacting one, is formally the same as the one in the KdV equation, however it is now valued on a non-associative and a non-alternative algebra. The other interacting term does not appear in the KdV equation and is similar to the interaction of a vector and a spinor field on a supersymmetric theory. It is a symmetry breaking term. For the octonion algebra the symmetry of the KdV equation reduces to the $SU(3)$ group, which is interesting since it is expected that the physics of the quarks can be modelled in terms of octonions.
 
 We obtained a B\"{a}cklund transformation in the sense of W-E for the non-linear integrable equation. It is a non-trivial transformation since it works for a general C-D algebra with a new interaction term, which is absent in the scalar  KdV equation. From the B\"{a}cklund transformation we got the Lax pair associated to the KdV equation valued on the C-D algebra. The extension of the corresponding eigenvalued problem can in principle be well formulated on a Hilbert space defined on a division algebra, that is up to the octonions but not beyond them.

 We also obtained the associated Gardner transformation for the non-linear equation. From it, by inverting the Gardner transformation we show the existence of an infinite sequence of conserved quantities for any C-D algebra. From the generalized Gardner equation we obtained the modified KdV equation on a C-D algebra. It also has an infinite sequence of conserved quantities.
 
 The new non-linear equation we proposed has additional symmetries besides its invariance under Galileo transformations. It is invariant under the automorphisms of the C-D algebra which preserve the  interaction source $v$. In particular for the octonion algebra it is invariant under the $SU(3)$ group.
 
 The B\"{a}cklund transformation besides its use in proving the existence of an infinite sequence of conserved quantities allows, in the usual way, to obtain solutions to the non-linear equation starting from the trivial one. From it we got the one-soliton and two-soliton solutions valued on the octonion algebra. They are the generalizations of the quaternion solutions found in \cite{Huang}.
 
 An interesting open problem is the existence of a bi-hamiltonian structure associated to all the integrable systems we have considered. For the KdV equation the first and second hamiltonian structures were obtained in \cite{Gardner10,Magri10}. Recently the bi-hamiltonian structures of KdV type for several integrable systems was obtained in \cite{Lorenzoni}.

 $\bigskip$

\textbf{Acknowledgments}

A. R. and A. S. are partially supported by Project Fondecyt
1161192, Chile.

\end{document}